\documentclass[11pt,preprintnumbers,titlepage]{revtex4-1}
\usepackage[latin9]{inputenc}
\usepackage{amsmath}
\usepackage{amssymb}
\usepackage{graphicx}
\usepackage[unicode=true,
 bookmarks=false,
 breaklinks=false,pdfborder={0 0 1},backref=section,colorlinks=false]
 {hyperref}
\hypersetup{
 colorlinks,linkcolor=blue,citecolor=blue,urlcolor=blue}

\makeatletter
\usepackage{amsfonts}
\usepackage{subfigure}
\usepackage{cleveref}
\usepackage{listings}
\usepackage{xcolor}
\usepackage{array}
\usepackage{url}
\usepackage{color}
\@ifundefined{textcolor}{}{
\definecolor{BLACK}{gray}{0}
\definecolor{WHITE}{gray}{1}
\definecolor{RED}{rgb}{1,0,0}
\definecolor{GREEN}{rgb}{0,1,0}
\definecolor{BLUE}{rgb}{0,0,1}
\definecolor{CYAN}{cmyk}{1,0,0,0}
\definecolor{MAGENTA}{cmyk}{0,1,0,0}
\definecolor{YELLOW}{cmyk}{0,0,1,0}
}

\makeatother

\begin{document}
\preprint{CTP-SCU/2022004}
\title{Echoes from Hairy Black Holes}
\author{Guangzhou Guo$^{a}$}
\email{gzguo@stu.scu.edu.cn}

\author{Peng Wang$^{a}$}
\email{pengw@scu.edu.cn}

\author{Houwen Wu$^{a,b}$}
\email{hw598@damtp.cam.ac.uk}

\author{Haitang Yang$^{a}$}
\email{hyanga@scu.edu.cn}

\affiliation{$^{a}$Center for Theoretical Physics, College of Physics, Sichuan
University, Chengdu, 610064, China}
\affiliation{$^{b}$Department of Applied Mathematics and Theoretical Physics,
University of Cambridge, Wilberforce Road, Cambridge, CB3 0WA, UK}
\begin{abstract}
We study the waveforms of time signals produced by scalar perturbations
in static hairy black holes, in which the perturbations can be governed
by a double-peak effective potential. The inner potential peak would
give rise to echoes, which provide a powerful tool to test the Kerr
hypothesis. The waveforms are constructed in the time\ and frequency
domains, and we find that the late-time waveforms are determined by
the long-lived and sub-long-lived quasinormal modes, which are trapped
in the potential valley and near the smaller peak, respectively. When
the distance between the peaks is significantly larger than the width
of the peaks, a train of decaying echo pulses is produced by the superposition
of the long-lived and sub-long-lived modes. In certain cases, the
echoes can vanish and then reappear. When the peaks are close enough,
one detects far fewer echo signals and a following sinusoid tail,
which is controlled by the long-lived or sub-long-lived mode and hence
decays very slowly.
\end{abstract}
\maketitle
\tableofcontents{}

\section{Introduction}

The existence of black holes in the universe is one of the most prominent
predictions of general relativity. Owing to advanced observation techniques
developed in the past decade, we are capable of exploring the nature
of black holes and testing general relativity in the strong field
regime. Gravitational waves from a binary black hole merger were successfully
detected by LIGO and Virgo \cite{Abbott:2016blz}, and subsequently
the first image of a supermassive black hole at the center of galaxy
M87 was photographed by the Event Horizon Telescope \cite{Akiyama:2019cqa,Akiyama:2019brx,Akiyama:2019sww,Akiyama:2019bqs,Akiyama:2019fyp,Akiyama:2019eap},
which opens a new era of black hole physics. Specifically, the ringdown
waveforms of gravitational waves are characterized by quasinormal
modes of final black holes \cite{Nollert:1999ji,Berti:2007dg,Cardoso:2016rao},
and hence the measurements of the ringdown waveforms offer new opportunities
to probe the detailed properties of black hole spacetime, e.g., the
black hole mass and spin \cite{Price:2017cjr,Giesler:2019uxc}.

Although current observations are found to be in good agreement with
the predictions of general relativity, observational uncertainties
still leave some room for alternatives to the Kerr black hole. In
particular, horizonless exotic compact objects (ECOs), e.g., boson
stars, gravastars and wormholes, have attracted a lot of attentions
\cite{Lemos:2008cv,Cunha:2017wao,Cunha:2018acu,Shaikh:2018oul,Dai:2019mse,Huang:2019arj,Simonetti:2020ivl,Wielgus:2020uqz,Yang:2021diz,Bambi:2021qfo,Peng:2021osd}.
Intriguingly, echo signals associated with the post-merger ringdown
phase in the binary black hole waveforms can be found in various ECO
models \cite{Cardoso:2016rao,Bueno:2017hyj,Konoplya:2018yrp,Wang:2018cum,Wang:2018mlp,Cardoso:2019rvt,GalvezGhersi:2019lag,Liu:2020qia,Ou:2021efv}.
Moreover, the recent LIGO/Virgo data may show the potential evidence
of echoes in gravitational wave waveforms of binary black hole mergers
\cite{Abedi:2016hgu,Abedi:2017isz}. As anticipated, echoes are closely
related to quasinormal modes \cite{Bueno:2017hyj,Price:2017cjr,Dey:2020pth}.
Specifically, it was argued that echoes are dominated by long-lived
quasinormal modes of ECOs, and the echo waveforms can be accurately
reconstructed from the quasinormal modes \cite{Bueno:2017hyj}. Moreover,
quasinormal modes of ECOs can be extracted from echo signals by a
Prony method \cite{Berti:2007dg,Yang:2021cvh}, which can be used
to approximately reconstruct effective potentials of the ECO spacetime
\cite{Volkel:2017kfj}. To gain a deeper insight into the generation
of echoes, a reflecting boundary was placed in a black hole spacetime
to mimic ECOs, and it showed that the reflecting boundary plays a
central role in producing extra time-delay echo pluses, which constitute
the echo waveform received by a distant observer \cite{Mark:2017dnq}.
This observation implies that echoes are expected to occur in the
spacetime with a double-peak effective scattering potential (e.g.,
wormholes), where the inner potential peak acts as a reflecting boundary.

Contrary to the common lore that detections of echoes in late-time
ringdown signals can be used to distinguish black holes with wormholes,
echo signals have been reported in several black hole models, e.g.,
quantum black holes \cite{Wang:2019rcf,Saraswat:2019npa,Dey:2020lhq,Oshita:2020dox,Chakraborty:2022zlq},
black holes with discontinuous effective potentials \cite{Liu:2021aqh},
nonuniform area quantization on black hole \cite{Chakravarti:2021clm}
and gravitons with modified dispersion relations \cite{DAmico:2019dnn}.
Remarkably, much less radical proposals for echoes from black holes
do exist in the literature. In fact, echoes have been found in dyonic
black holes with a quasi-topological electromagnetic term, which have
multiple photon spheres and double-peak effective potentials \cite{Liu:2019rib,Huang:2021qwe}.
In a black hole of massive gravity, gravitational perturbations should
couple with the background metric and Stuckelberg fields, which could
give echo signals in the gravitational waves \cite{deRham:2010kj,Dong:2020odp}.
Given the theoretical and observational importance of echoes, it is
of great significance to find more black hole spacetimes that can
produce echo signals.

Recently, a novel type of hairy black hole solutions were constructed
in Einstein-Maxwell-scalar (EMS) models \cite{Herdeiro:2018wub,Konoplya:2019goy,Wang:2020ohb,Guo:2021zed,Guo:2021ere},
which serve as counter-examples to the no-hair theorem \cite{Israel:1967wq,Carter:1971zc,Ruffini:1971bza}.
In the EMS models, the scalar field is non-minimally coupled to the
electromagnetic field and can trigger a tachyonic instability to form
spontaneously scalarized hairy black holes from Reissner-Nordström
(RN) black holes. Properties of the hairy black holes have been extensively
studied in the literature, e.g., different non-minimal coupling functions
\cite{Fernandes:2019rez,Fernandes:2019kmh,Blazquez-Salcedo:2020nhs},
massive and self-interacting scalar fields \cite{Zou:2019bpt,Fernandes:2020gay},
horizonless reflecting stars \cite{Peng:2019cmm}, stability analysis
of hairy black holes \cite{Myung:2018vug,Myung:2019oua,Zou:2020zxq,Myung:2020etf,Mai:2020sac},
higher dimensional scalar-tensor models \cite{Astefanesei:2020qxk},
quasinormal modes of hairy black holes \cite{Myung:2018jvi,Blazquez-Salcedo:2020jee},
two U$\left(1\right)$ fields \cite{Myung:2020dqt}, quasi-topological
electromagnetism \cite{Myung:2020ctt}, topology and spacetime structure
influences \cite{Guo:2020zqm}, and scalarized black holes in the
dS/AdS spacetime \cite{Brihaye:2019dck,Brihaye:2019gla,Zhang:2021etr,Guo:2021zed}.

In recent works \cite{Gan:2021pwu,Gan:2021xdl}, we found that the
hairy black holes can also possess multiple photon spheres outside
the event horizon, which have significant effects on the optical observation
of black holes illuminated by the surrounding accretion disk, e.g.,
leading to bright rings of different radii in the black hole images
\cite{Gan:2021pwu} and significantly increasing the flux of the observed
images \cite{Gan:2021xdl}. Later, it showed that the effective potential
for a scalar perturbation in the hairy black holes exhibits a double-peak
structure \cite{Guo:2021enm}. In the eikonal limit, the extrema of
the double-peak potential correspond to the photon spheres, around
which long-lived and sub-long-lived quasinormal modes were found \cite{Guo:2021enm}.
The appearance of the double-peak effective potentials naturally motivates
us to search echo signals of perturbations in the hairy black holes.
Moreover, it is highly desirable to explore the relationship between
echoes and the long-lived and sub-long-lived modes obtained in \cite{Guo:2021enm}.
To this end, we numerically obtain time-domain echoes of scalar perturbations
in the hairy black holes and reconstruct the late-time signals from
associated quasinormal modes in this paper. The remainder of this
paper is organized as follows. In Section \ref{SetUp}, after the
hairy black hole solution is briefly reviewed, we consider a time-dependent
scalar field perturbation propagating in the hairy black holes and
relate it to the quasinormal mode spectrum. The evolutions of the
scalar field perturbation in different profiles of effective potentials
are exhibited in Section \ref{sec:Numerical-Result}. We finally conclude
our main results in Section \ref{sec:Conclusions}. We set $16\pi G=1$
throughout this paper.

\section{Set Up}

\label{SetUp}

In this section, we first briefly review spherically symmetric hairy
black hole solutions in the EMS model. In the hairy black hole background,
the evolution of time-dependent scalar perturbations from initial
data is then studied and related to the corresponding quasinormal
modes.

\subsection{Hairy Black Holes}

The scalar field $\phi$ is minimally coupled to the metric field
and non-minimally coupled to the electromagnetic field $A_{\mu}$
in the EMS model, which is described by the action 
\begin{equation}
S=\int d^{4}x\sqrt{-g}\left[R-2\left(\partial\phi\right)^{2}-e^{\alpha\phi^{2}}F^{2}\right].\label{eq:Action}
\end{equation}
Here, $F_{\mu\nu}=\partial_{\mu}A_{\nu}-\partial_{\nu}A_{\mu}$ is
the electromagnetic field strength tensor, and $e^{\alpha\phi^{2}}$
is the coupling function between $\phi$ and $A_{\mu}$. With the
spherically symmetric and asymptotically flat black hole ansatz \cite{Herdeiro:2018wub,Guo:2021zed},
\begin{align}
ds^{2} & =-N\left(r\right)e^{-2\delta\left(r\right)}dt^{2}+\frac{1}{N\left(r\right)}dr^{2}+r^{2}\left(d\theta^{2}+\sin^{2}\theta d\varphi^{2}\right),\nonumber \\
A_{\mu}dx^{\mu} & =V\left(r\right)dt\text{ and }\phi=\phi\left(r\right),\label{eq:ansatz}
\end{align}
the equations of motion for the action $\left(\ref{eq:Action}\right)$
are given by 
\begin{align}
m^{\prime}\left(r\right) & =\frac{Q^{2}}{2r^{2}e^{\alpha\phi^{2}\left(r\right)}}+\frac{1}{2}r^{2}N\left(r\right)\phi^{\prime2}\left(r\right)\nonumber \\
\left[r^{2}N\left(r\right)\phi^{\prime}\left(r\right)\right]^{\prime} & =-\frac{\alpha\phi\left(r\right)Q^{2}}{e^{\alpha\phi^{2}\left(r\right)}r^{2}}-r^{3}N\left(r\right)\phi^{\prime3}\left(r\right),\nonumber \\
\delta^{\prime}\left(r\right) & =-r\phi^{\prime2}\left(r\right),\label{eq:NLEqs}\\
V^{\prime}\left(r\right) & =\frac{Q}{r^{2}e^{\alpha\phi^{2}\left(r\right)}}e^{-\delta\left(r\right)},\nonumber 
\end{align}
where primes denote derivatives with respect to $r$. In the above
equations $\left(\ref{eq:NLEqs}\right)$, the integration constant
$Q$ is interpreted as the electric charge of the black hole solution,
and the Misner-Sharp mass function $m\left(r\right)$ is defined via
$N\left(r\right)\equiv1-2m\left(r\right)/r$.

\begin{figure}[ptb]
\begin{centering}
\includegraphics[scale=0.77]{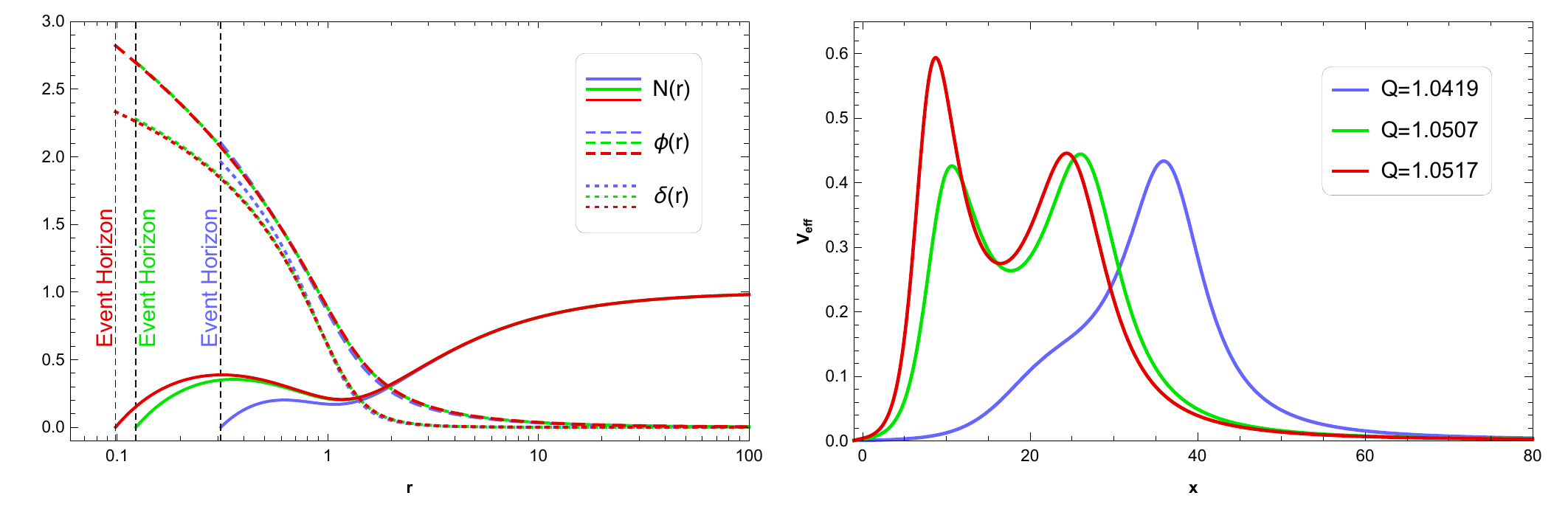} 
\par\end{centering}
\caption{Hairy black hole solutions for $Q=1.0419$ (blue lines), $Q=1.0507$
(green lines) and $Q=1.0517$ (red lines) with $\alpha=0.8$ and the
corresponding $l=2$ effective potentials of the scalar field. \textbf{Left}:
The metric functions are plotted outside the event horizon (vertical
dashed lines), and the solid, dashed and dotted lines designate $N\left(r\right)$,
$\phi\left(r\right)$ and $\delta\left(r\right)$, respectively. \textbf{Right}:
The effective potentials are displayed against the tortoise coordinate
$x$. For a small value of the charge (e.g., the blue line), the effective
potential has only a single extremum. As the black hole charge increases
(e.g., the green and red lines), the effective potential presents
a double-peak structure with two local maxima and one local minimum.}
\label{Hbh-plot}
\end{figure}

Black hole solutions of the non-linear ordinary differential equations
$\left(\ref{eq:NLEqs}\right)$ can be obtained as appropriate boundary
conditions at the event horizon $r_{h}$ and the spatial infinity
are imposed, 
\begin{align}
m(r_{h}) & =r_{h}/2\text{, }\delta(r_{h})=\delta_{0}\text{, }\phi(r_{h})=\phi_{0}\text{, }V(r_{h})=0\text{,}\nonumber \\
m(\infty) & =M\text{, }\delta(\infty)=0\text{, }\phi(\infty)=0\text{, }V(\infty)=\Phi\text{,}\label{eq:BC}
\end{align}
where $\Phi$ is the electrostatic potential, and the black hole mass
$M$ is related to the ADM mass. The free parameters $\delta_{0}$
and $\phi_{0}$ can be used to characterize different black hole solutions.
Specifically, $\phi_{0}=\delta_{0}=0$ lead to the scalar-free solutions
with $\phi=0$ of eqn. $\left(\ref{eq:NLEqs}\right)$, which are exactly
RN black holes. Nevertheless, hairy black hole solutions with a non-trivial
scalar field $\phi$ can exist if non-zero values of $\phi_{0}$ and
$\delta_{0}$ are admitted. In this paper, we set $M=1$ and use a
shooting method built in the \textit{$NDSolve$} function of \textit{$Wolfram\text{ }\circledR Mathematica$}
to numerically solve eqn. $\left(\ref{eq:NLEqs}\right)$ with the
given boundary conditions $\left(\ref{eq:BC}\right)$. The metric
functions of three hairy black hole solutions with $\alpha=0.8$ are
exhibited in FIG. \ref{Hbh-plot}, where the blue, green and red lines
denote $Q=1.0419$, $Q=1.0507$ and $Q=1.0517$, respectively.

\subsection{Time-dependent Scalar Field Perturbations}

For a scalar field perturbation $\delta\phi$ around the hairy black
hole, the master equation is given by \cite{Herdeiro:2018wub,Guo:2021enm}
\begin{equation}
\left[\nabla^{\mu}\nabla_{\mu}+\left(\alpha+2\alpha^{2}\phi^{2}\left(r\right)\right)\frac{Q^{2}}{r^{4}e^{\alpha\phi^{2}\left(r\right)}}\right]\delta\phi=0.\label{eq:master eq}
\end{equation}
For later convenience, we define the tortoise coordinate $x$ via
$dx/dr\equiv e^{\delta\left(r\right)}N^{-1}\left(r\right)$. The time-dependent
scalar field perturbation $\delta\phi$ can be decomposed in terms
of spherical harmonics, 
\begin{equation}
\delta\phi=\sum_{l,m}\frac{\psi\left(t,r\right)}{r}Y_{lm}\left(\theta,\varphi\right).\label{eq:psi-rt}
\end{equation}
With the help of eqns. $\left(\ref{eq:ansatz}\right)$ and $\left(\ref{eq:psi-rt}\right)$,
the master equation $\left(\ref{eq:master eq}\right)$ then reduces
to 
\begin{equation}
\left(-\frac{\partial^{2}}{\partial t^{2}}+\frac{\partial^{2}}{\partial x^{2}}-V_{\text{eff}}\left(x\right)\right)\psi\left(t,x\right)=0,\label{eq:t-x eq}
\end{equation}
where the effective potential $V_{\text{eff}}\left(x\right)$ is given
by 
\begin{equation}
V_{\text{eff}}\left(x\right)=\frac{e^{-2\delta\left(r\right)}N\left(r\right)}{r^{2}}\left[l\left(l+1\right)+1-N\left(r\right)-\frac{Q^{2}}{r^{2}e^{\alpha\phi\left(r\right)^{2}}}-\left(\alpha+2\alpha^{2}\phi\left(r\right)^{2}\right)\frac{Q^{2}}{r^{2}e^{\alpha\phi\left(r\right)^{2}}}\right].\label{eq:Veff}
\end{equation}
The effective potential with $l=2$ of various black hole solutions
is presented in the right panel of FIG. \ref{Hbh-plot}. Intriguingly,
when the black hole charge is large enough, the effective potential
can possess a double-peak structure, which consists of two local maxima
and one local minimum.

To solve the partial differential equation $\left(\ref{eq:t-x eq}\right)$,
we consider a time-dependent Green's function $G\left(t,x,x^{\prime}\right)$,
which satisfies 
\begin{equation}
\left(-\frac{\partial^{2}}{\partial t^{2}}+\frac{\partial^{2}}{\partial x^{2}}-V_{\text{eff}}\left(x\right)\right)G\left(t,x,x^{\prime}\right)=\delta\left(t\right)\delta\left(x-x^{\prime}\right).
\end{equation}
One then can express the solution of eqn. $\left(\ref{eq:t-x eq}\right)$
in terms of $G\left(t,x,x^{\prime}\right)$ \cite{Andersson:1996cm,Nollert:1999ji},
\begin{equation}
\psi\left(t,x\right)=-\int_{-\infty}^{+\infty}\left[G\left(t,x,x^{\prime}\right)\partial_{t}\psi\left(0,x^{\prime}\right)+\partial_{t}G\left(t,x,x^{\prime}\right)\psi\left(0,x^{\prime}\right)\right]dx^{\prime}.\label{eq:psi(t,x)-0}
\end{equation}
Under the Fourier transformation, the solution $\left(\ref{eq:psi(t,x)-0}\right)$
can be rewritten as 
\begin{equation}
\psi\left(t,x\right)=\frac{1}{2\pi}\int_{-\infty}^{+\infty}\hat{G}\left(\omega,x,x^{\prime}\right)\hat{S}\left(\omega,x\right)e^{-i\omega t}dx^{\prime}d\omega,\label{eq:psi(t,x)-1}
\end{equation}
where $\hat{S}\left(\omega,x\right)$ is determined by the initial
data, 
\begin{equation}
\hat{S}\left(\omega,x\right)=\left.\left[i\omega\psi\left(t,x\right)-\frac{\partial\psi\left(t,x\right)}{\partial t}\right]\right\vert _{t=0}.\label{eq:Initial Data}
\end{equation}

The time-independent Green's function $\hat{G}\left(\omega,x,x^{\prime}\right)$
can be constructed in terms of two linearly independent solutions
$\hat{\psi}_{-}\left(\omega,x\right)$ and $\hat{\psi}_{+}\left(\omega,x\right)$
to the homogeneous differential equation, 
\begin{equation}
\left(\frac{\partial^{2}}{\partial^{2}x}+\omega^{2}-V_{\text{eff}}\left(x\right)\right)\hat{\psi}\left(\omega,x\right)=0,\label{eq:x-w eq}
\end{equation}
with the boundary conditions 
\begin{align}
\hat{\psi}_{-}\left(\omega,x\right) & \sim e^{-i\omega x},\qquad x\rightarrow-\infty,\nonumber \\
\hat{\psi}_{+}\left(\omega,x\right) & \sim e^{i\omega x},\qquad x\rightarrow+\infty.\label{eq:BC+-}
\end{align}
Particularly, the Green's function $\hat{G}\left(\omega,x,x^{\prime}\right)$
is given by 
\begin{equation}
\hat{G}\left(\omega,x,x^{\prime}\right)=\frac{\hat{\psi}_{-}\left(\omega,\min\left(x,x^{\prime}\right)\right)\hat{\psi}_{+}\left(\omega,\max\left(x,x^{\prime}\right)\right)}{W\left(\omega\right)},
\end{equation}
where the Wronskian $W\left(\omega\right)$ is defined as 
\begin{equation}
W\left(\omega\right)=\hat{\psi}_{-}\left(\omega,x\right)\partial_{x}\hat{\psi}_{+}\left(\omega,x\right)-\partial_{x}\hat{\psi}_{-}\left(\omega,x\right)\hat{\psi}_{+}\left(\omega,x\right).
\end{equation}
When $W\left(\omega\right)=0$, one can infer that $\hat{\psi}_{-}\left(\omega,x\right)$
is identical to $\hat{\psi}_{+}\left(\omega,x\right)$ up to a constant
factor, which indicates that $\hat{\psi}_{\pm}\left(\omega,x\right)$
have only ingoing (outgoing)\ modes at $x=-\infty$ ($x=+\infty$).
Therefore, the condition $W\left(\omega\right)=0$ selects a discrete
set of quasinormal modes with complex quasinormal frequencies $\omega_{n}$,
where $n=0,1,2\ldots$ is the overtone number.

On the complex plane, the solution $\psi\left(t,x\right)$ to eqn.
$\left(\ref{eq:t-x eq}\right)$ can be expressed as a sum of quasinormal
modes \cite{Nollert:1999ji}, 
\begin{equation}
\psi\left(t,x\right)=\sum_{n}\psi_{n}\left(t,x\right)=\sum_{n}c_{n}\left(x\right)e^{-i\omega_{n}t},\label{eq:psi of QNMs}
\end{equation}
where the coefficient $c_{n}\left(x\right)$ is 
\begin{equation}
c_{n}\left(x\right)=\frac{-i}{dW\left(\omega_{n}\right)/d\omega}\int_{-\infty}^{+\infty}\hat{\psi}_{-}\left(\omega_{n},\min\left(x,x^{\prime}\right)\right)\hat{\psi}_{+}\left(\omega_{n},\max\left(x,x^{\prime}\right)\right)\hat{S}\left(\omega_{n},x^{\prime}\right)dx^{\prime}.\label{eq:cn}
\end{equation}
Since the quasinormal modes come in complex conjugate pairs, the waveform
$\psi\left(t,x\right)$ given by eqn. $\left(\ref{eq:psi of QNMs}\right)$
is real as long as the initial data $\hat{S}\left(\omega_{n},x\right)$
is real \cite{Nollert:1999ji}. Note that, before the initial data
is entirely received by the observer, a time-dependent integration
domain in eqns. $\left(\ref{eq:psi(t,x)-0}\right)$ and $\left(\ref{eq:cn}\right)$
is required to respect causality. Therefore, to well describe the
behavior of $\psi\left(t,x\right)$ at a early time, the coefficients
$c_{n}\left(x\right)$ are argued to depend on time \cite{Andersson:1996cm,Nollert:1999ji}.
Nevertheless, we focus on the late-time waveforms throughout this
paper, and hence the coefficients $c_{n}\left(x\right)$ are time-independent.

When the real parts of quasinormal frequencies $\omega_{n}$ are an
arithmetic progression with regard to the number $n$, 
\begin{equation}
\operatorname{Re}\omega_{n}=\omega_{0}+\frac{2n\pi}{T},\label{eq:AP modes}
\end{equation}
the waveform $\psi\left(t,x\right)$ from eqn. $\left(\ref{eq:psi of QNMs}\right)$
with real initial data then behaves as 
\begin{equation}
\psi\left(t,x\right)=\sum_{n}\psi_{n}\left(t,x\right)=2\sum_{n}\left[\operatorname{Re}c_{n}\cos\left(\omega_{0}+2n\pi/T\right)t+\operatorname{Im}c_{n}\sin\left(\omega_{0}+2n\pi/T\right)t\right]e^{\operatorname{Im}\omega_{n}t},\label{eq:SinCos psi}
\end{equation}
where each quasinormal mode $\psi_{n}\left(t,x\right)$ is composed
of a complex conjugate pair, 
\begin{equation}
\psi_{n}\left(t,x\right)=c_{n}\left(x\right)e^{-i\omega_{n}t}+c_{n}^{\ast}\left(x\right)e^{-i(-\omega_{n}^{\ast})t}.
\end{equation}
Interestingly, the waveform $\psi\left(t,x\right)$ describes damped
oscillations with a period $T$ and damping factors, which are the
imaginary parts of quasinormal modes. As demonstrated in \cite{Bueno:2017hyj},
the inner barrier of a double-peak potential provides a reflecting
wall for radiation waves, leading to a set of quasinormal modes in
the form of eqn. $\left(\ref{eq:AP modes}\right)$. Consequently,
a distant observer can detect a series of echoes from a double-peak
effective potential.

\section{Numerical Results}

\label{sec:Numerical-Result}

In this section, we investigate the waveform $\psi\left(t,x\right)$
detected by a distant observer in the hairy black holes with the effective
potential of different peak structure. To numerically solve the partial
differential equation $\left(\ref{eq:t-x eq}\right)$, we consider
the initial condition, 
\begin{equation}
\left.\psi\left(t,x\right)\right\vert _{t=0}=0\text{ and}\left.\frac{\partial\psi\left(t,x\right)}{\partial t}\right\vert _{t=0}=A\text{exp}\left(-\frac{\left(x-x_{0}\right)^{2}}{2\Delta^{2}}\right)\text{ \cite{Cardoso:2016oxy,Cardoso:2017njb,Cardoso:2019rvt}}.\label{eq:initial data}
\end{equation}
In the following numerical simulations, the initial position of the
Gaussian wave packet $x_{0}$ is placed near the (outer) peak of the
effective potential, and the amplitude $A$ and the width $\Delta$
are chosen to adapt to the specific case. To check our numerical results
and find the frequency content of $\psi\left(t,x\right)$, we use
eqns. $\left(\ref{eq:psi of QNMs}\right)$ and $\left(\ref{eq:cn}\right)$
to reconstruct the waveform $\psi\left(t,x\right)$ at late times
from the quasinormal modes. Moreover, we focus on the spherical harmonics
of $l=2$ since they play a dominant role in the ringdown gravitational
waves after binary black holes merge \cite{Giesler:2019uxc,Sago:2021gbq}.

\subsection{Single-peak Potential}

\begin{figure}[ptb]
\begin{centering}
\includegraphics[scale=0.72]{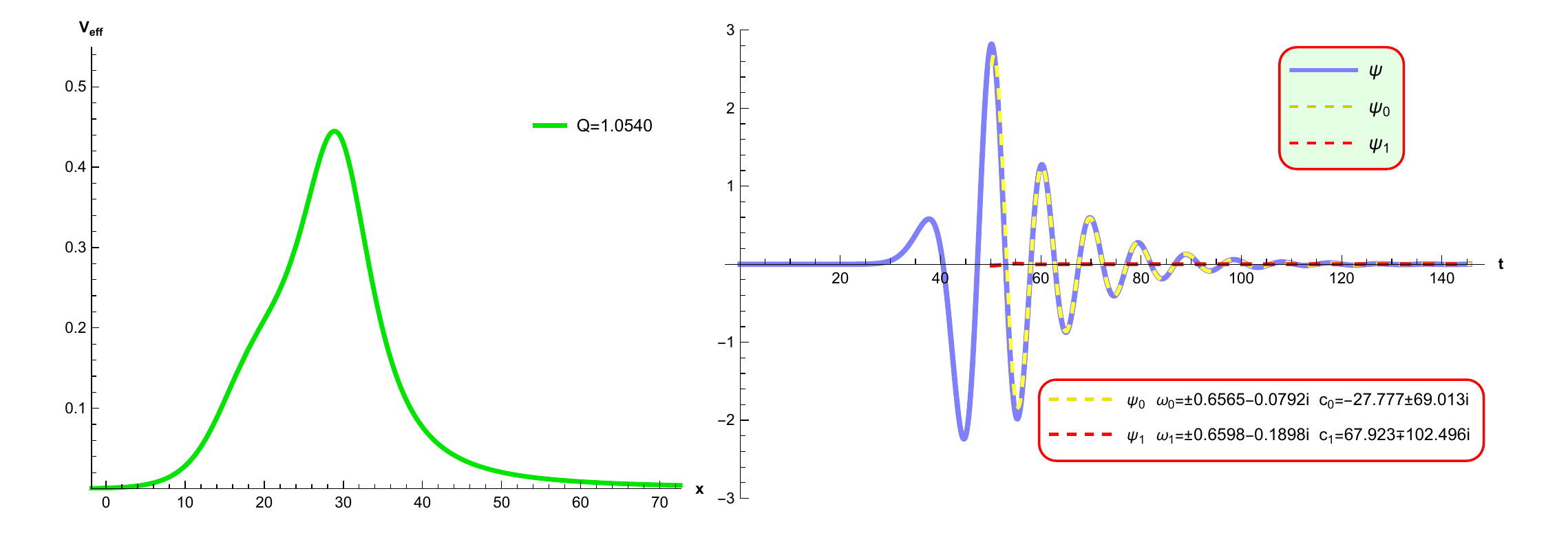} 
\par\end{centering}
\caption{\textbf{Left}: The $l=2$ effective potential of the hairy black hole
with $\alpha=0.9$ and $Q=1.0540$ as a function of the tortoise coordinate
$x$. The effective potential has a single-peak structure. \textbf{Right}:
The waveform $\psi$ (blue solid line), and the first and second lowest-lying
quasinormal modes $\psi_{0}$ (yellow dashed line) and $\psi_{1}$
(red dashed line) received by an observer at $x_{\text{o}}=72.75$.
The initial condition is the Gaussian perturbation $\left(\ref{eq:initial data}\right)$
near the peak. The fundamental mode $\psi_{0}$ dominates the late-time
behavior of the waveform $\psi$. }
\label{alpha09}
\end{figure}

As shown in FIG. $\ref{Hbh-plot}$, the $l=2$ effective potential
of scalar field perturbations has a single-peak structure when the
black hole charge is small enough. In FIG. $\ref{alpha09}$, we present
the evolution of a time-dependent scalar perturbation in the hairy
black hole with $\alpha=0.9$ and $Q=1.0540$. The $l=2$ effective
potential is plotted in the left panel, which indeed shows a single-peak
structure. In the right panel, we display the waveform signal received
by an observer at $x_{\text{o}}=72.75$, who is far away from the
potential peak. Specifically, the blue solid line denotes the solution
$\psi(t,x_{\text{o}})$ to the partial differential equation $\left(\ref{eq:t-x eq}\right)$
with the initial condition $\left(\ref{eq:initial data}\right)$,
and the dashed lines represent the low-lying quasinormal modes $\psi_{0}(t,x_{\text{o}})$
and $\psi_{1}(t,x_{\text{o}})$ obtained from eqns. $\left(\ref{eq:psi of QNMs}\right)$,
$\left(\ref{eq:cn}\right)$ and $\left(\ref{eq:initial data}\right)$.
Here, we consider the quasinormal modes $\psi_{0}(t,x_{\text{o}})$
and $\psi_{1}(t,x_{\text{o}})$ only after the initial data is fully
received by the observer. The right panel of FIG. $\ref{alpha09}$
displays that, roughly after the travel time of the initial data from
the vicinity of the peak to the observer, the reflection from the
potential peak gives an observed burst, which can be accurately reconstructed
from $\psi_{0}$ and $\psi_{1}$. At late times, the wave signal is
dominated by the fundamental quasinormal mode $\psi_{0}$, showing
an exponentially damped sinusoid. As expected, due to the absence
of the inner peak, no echo is observed after the burst is received.
Note that waves propagating on a black hole spacetime usually develop
asymptotically late-time tails, which follow exponentially damped
sinusoids and decay as an inverse power of time due to scattering
from large radius in the black hole geometry \cite{Price:1972pw,Ching:1994bd}.
Nevertheless, discussions on the power-law tails are beyond the scope
of the paper.

\subsection{Wormhole-like Potential}

For a large enough black hole charge, there can exist two peaks in
the effective potential of scalar field perturbations. Depending on
the black hole parameters, the separation between the peaks $L$ can
be considerably larger than the Compton wavelength of the scalar field
perturbations (or the wavelength of the associated quasinormal modes),
which resembles the usual wormhole spacetime.\ Since the potential
peaks are well separated, the scattering of a perturbation off one
peak is barely influenced by the other one. The perturbation is reflected
off the potential barriers, and bounces back and forth between the
two peaks. Meanwhile, the perturbation successively tunnels through\ the
outer barrier, leading to a series of echoes received by a distant
observer. In this case, the geometric optics approximation is valid,
and hence the time delay between the echoes is roughly $2L$. For
a detail discussion, one can refer to \cite{Mark:2017dnq,Bueno:2017hyj}.
As noted in \cite{Bueno:2017hyj}, the appearance of echoes in a double-peak
potential is closely related to quasinormal modes residing in the
valley between the peaks. When the distance between the peaks is much
larger than the width of the peaks, there exists a series of these
quasinormal modes, whose imaginary parts are much smaller than their
real parts. Similar to a beat produced by multiple sounds of slightly
different frequencies, the superposition of the quasinormal modes,
which leak through the outer potential barrier, produce approximately
periodical echo signals.\
 
\begin{figure}[ptb]
\begin{centering}
\includegraphics[scale=0.7]{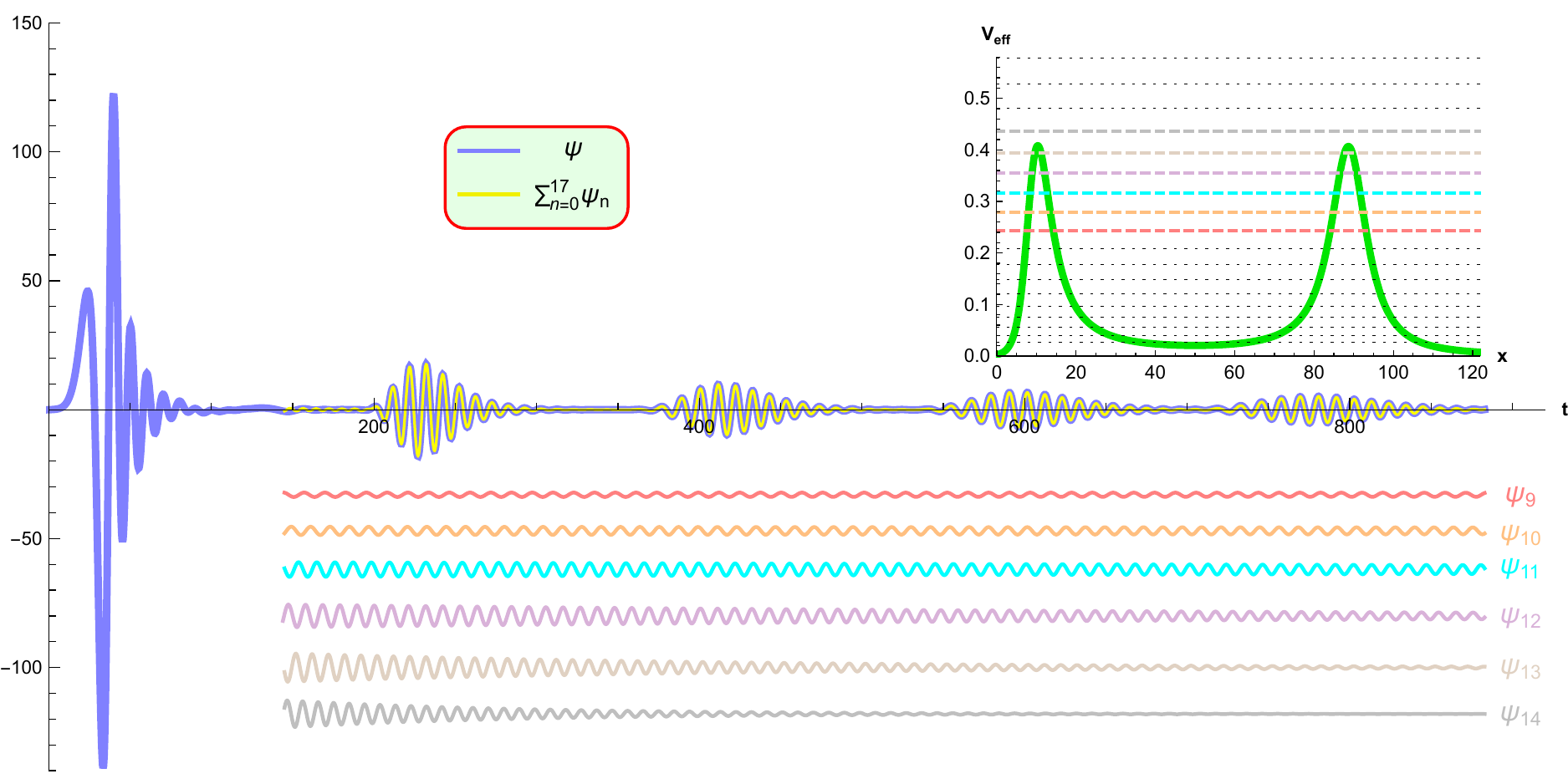} 
\par\end{centering}
\caption{The upper-right inset shows the $l=2$ effective potential of two
well-separated peaks\ in the hairy black hole with $\alpha=0.52$
and $Q=1.0074$. The blue line denotes the observed time signal, which
is obtained from numerical integration of eqn. $\left(\ref{eq:t-x eq}\right)$
with an initial Gaussian's packet near the outer peak. We take the
observer to be sufficiently far away from the outer peak. Due to the
double-peak structure, a sequence of echoes starts to appear after
the primary signal, and can be well reconstructed from the quasinormal
modes from $n=0$ to $n=17$ (yellow line). Below the time axis, we
exhibit the waveform of the quasinormal modes from $n=9$ to $n=14$
individually, which dominate the reconstruction of the echoes. Additionally,
$\left(\operatorname{Re}\omega_{n}\right)^{2}$ of the quasinormal
modes are plotted as horizontal lines in the inset. The smaller $n$
is, the lower the horizontal line lies. }
\label{alpha05}
\end{figure}

In FIG. $\ref{alpha05}$, we present the numerically computed and
reconstructed time signals received by an observer far away from the
outer peak of the effective potential in the hairy black hole with
$\alpha=0.52$ and $Q=1.0074$. The blue line designates the numerical
solution $\psi\left(t,x\right)$ to the partial differential equation
$\left(\ref{eq:t-x eq}\right)$, while the yellow line represents
the sum over the associated quasinormal modes from $n=0$ to $n=17$.
One can observe that a series of echoes roughly separated by a distance
$2L\approx$ $156.68$ arises at late times, and the sum of quasinormal
modes perfectly reconstruct the echoes. In TABLE \ref{alpha05n15},
we list the quasinormal frequency $\omega_{n}$ and the corresponding
coefficient $c_{n}$ for each quasinormal mode $\psi_{n}$. Roughly
speaking, these quasinormal frequencies $\omega_{n}$ satisfy the
form in eqn. $\left(\ref{eq:AP modes}\right)$ with a period $T\approx179.52\sim2L$,
which is consistent with the echo period. To illustrate the contributions
from quasinormal modes, the dominant modes $\psi_{9}\sim\psi_{14}$
are individually exhibited below the time axis of FIG. $\ref{alpha05}$.
Moreover, the squares of the real parts of the quasinormal frequencies
$\left(\operatorname{Re}\omega_{n}\right)^{2}$ are displayed as horizontal
lines in the upper-right inset. The dashed lines represent the dominant
modes with the same colors as those below the time axis, while the
dotted lines denote other quasinormal modes. The quasinormal modes
spread out beyond the potential valley by penetrating the potential
barriers. It shows that, the smaller $n$ is, the closer the quasinormal
mode lives to the bottom of the valley, thus making it penetrate the
potential barriers more difficult. Note that the coefficients $c_{n}$
are strongly related to the transmission of the scalar perturbation
penetrating the outer barrier, which implies that low-lying quasinormal
modes have small values of $c_{n}$. For the quasinormal modes $\psi_{n<9}$,
the coefficients $c_{n<9}$ are so small that their contributions
to the echoes can be neglected. On the other hand, the frequency of
the high-lying modes $\psi_{n>14}$ attains a large negative imaginary
part, providing a prominent exponentially damping factor for the time
signal.

\begin{table}[ptb]
\begin{centering}
\includegraphics[scale=1.02]{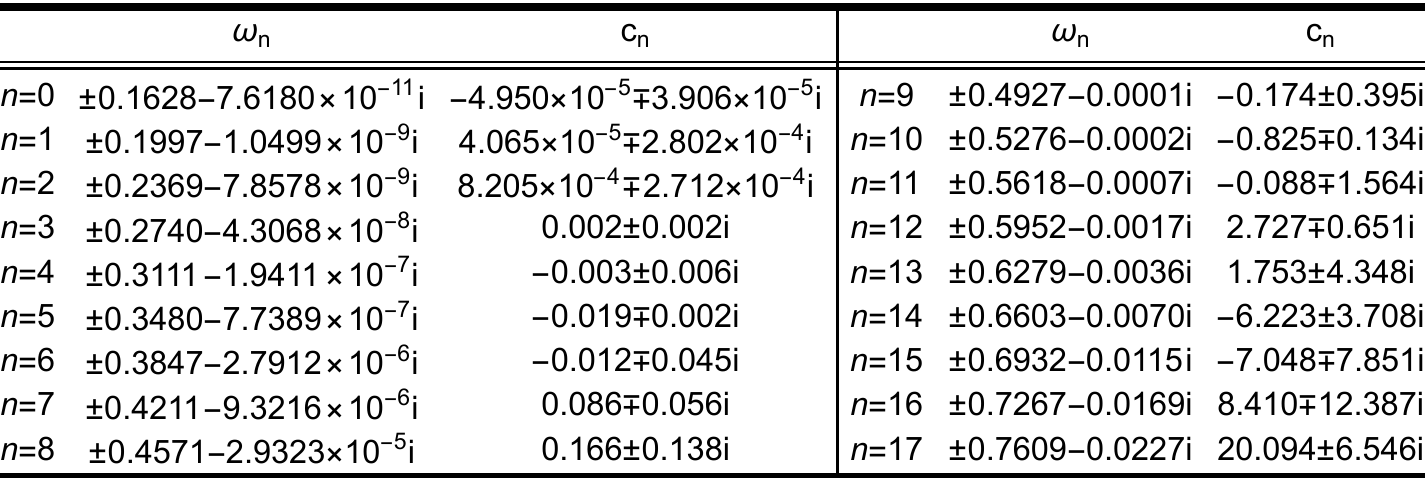} 
\par\end{centering}
\caption{The frequency $\omega_{n}$ of the quasinormal modes of $l=2$ for
a scalar field perturbation in the hairy black hole with $\alpha=0.52$
and $Q=1.0074$. The coefficient $c_{n}$, which controls the contribution
from the corresponding quasinormal mode to the echoes in FIG. $\ref{alpha05}$,
is obtained from eqn. $\left(\ref{eq:cn}\right)$. As $n$ grows,
the modulus of $c_{n}$ becomes larger. Since the initial data is
real, each pair of coefficients $c_{n}$ are complex conjugate, leading
to the real waveform of the quasinormal modes $\psi_{n}$.}
\label{alpha05n15}
\end{table}

\begin{figure}[ptb]
\begin{centering}
\includegraphics[scale=0.7]{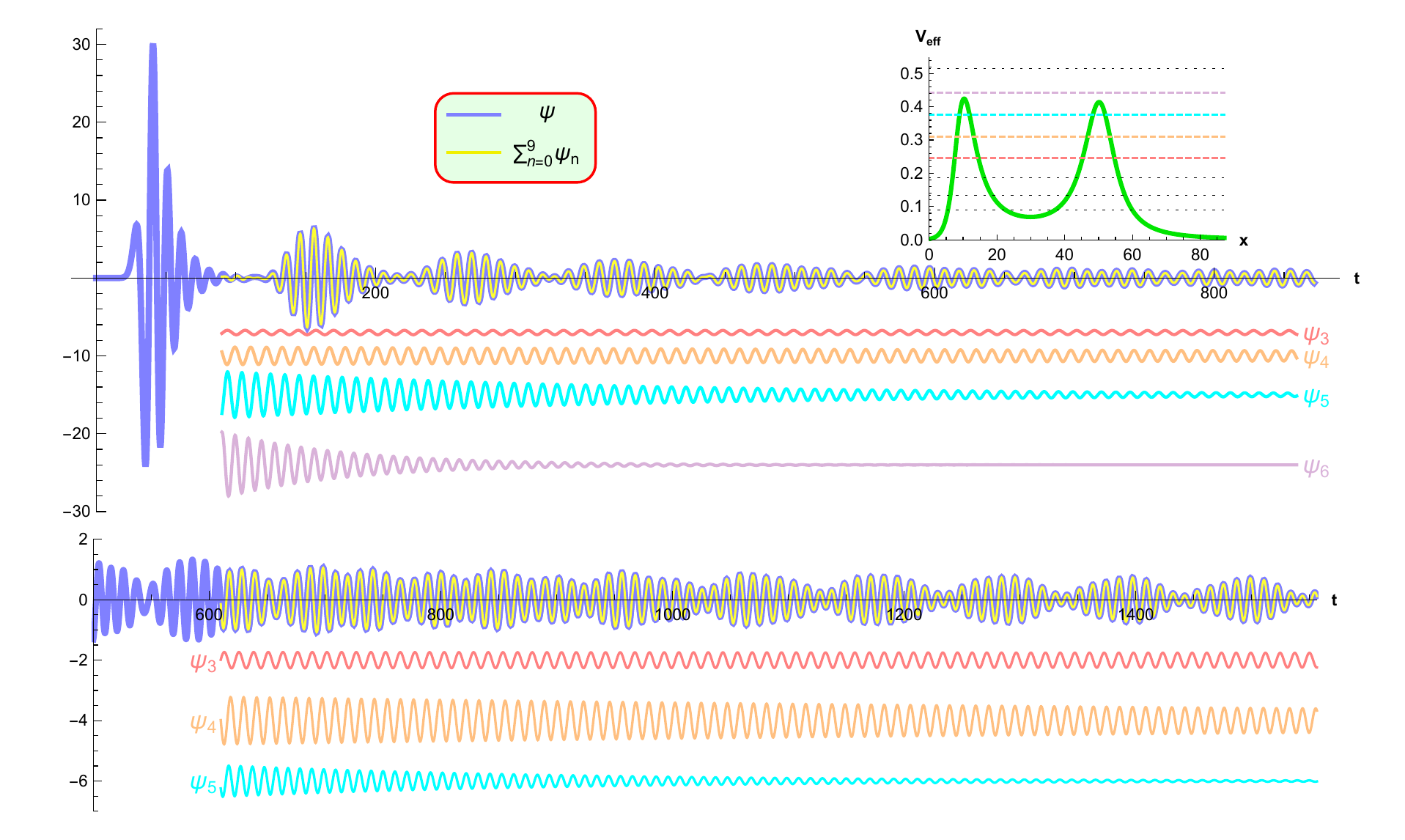} 
\par\end{centering}
\caption{Exact numerical (blue lines) and reconstructed (yellow lines) waveforms
$\psi$, and the dominant quasinormal modes received by a distant
observer in the hairy black hole with $\alpha=0.6$ and $Q=1.0192$.
Compared to FIG. $\ref{alpha05}$, the effective potential has a smaller
valley between two peaks, leading to fewer quasinormal modes (see
horizontal lines, representing $\left(\operatorname{Re}\omega_{n}\right)^{2}$
of quasinormal modes, in the inset). In the upper panel, the echoes
are observed after the primary signal and disappear after some time.
The lower panel shows the later behavior of the waveform. Interestingly,
the echoes reoccur after the quasinormal mode $\psi_{5}$ is damped
away. Specifically, the superposition of the quasinormal modes $\psi_{3}$
and $\psi_{4}$ produce the reoccurring echoes at the later time. }
\label{alpha06}
\end{figure}

In FIG. $\ref{alpha06}$, we display the numerically computed and
reconstructed waveforms at a position far away from the outer peak
in the hairy black hole solution with $\alpha=0.6$ and $Q=1.0192$,
where the effective potential has a smaller valley than that of FIG.
$\ref{alpha05}$. In addition, the dominant quasinormal modes of the
waveform $\psi$ are also exhibited below the time axis. Unlike FIG.
$\ref{alpha05}$, we observe that there exists a time regime (roughly
between $700$ and $1000$), in which echoes overlaps, and echo signals
can be hardly identified. Interestingly, after the quasinormal mode
$\psi_{5}$ becomes negligible, the echoes composed of the low-lying
modes $\psi_{3}$ and $\psi_{4}$ reappear. Moreover, the frequency
$\omega_{n}$ and the coefficient $c_{n}$ of the quasinormal modes
$\psi_{n\leq9}$ are given in TABLE \ref{alpha06n9}. Compared to
FIG. $\ref{alpha05}$ and TABLE \ref{alpha05n15}, we find that fewer
quasinormal modes are available to reconstruct the waveform due to
a smaller potential valley \cite{Guo:2021enm}. In fact, as shown
in the inset of FIG. $\ref{alpha06}$, the quasinormal modes get farther
apart as the potential valley becomes narrower and shallower.

\begin{table}[ptb]
\begin{centering}
\includegraphics[scale=0.95]{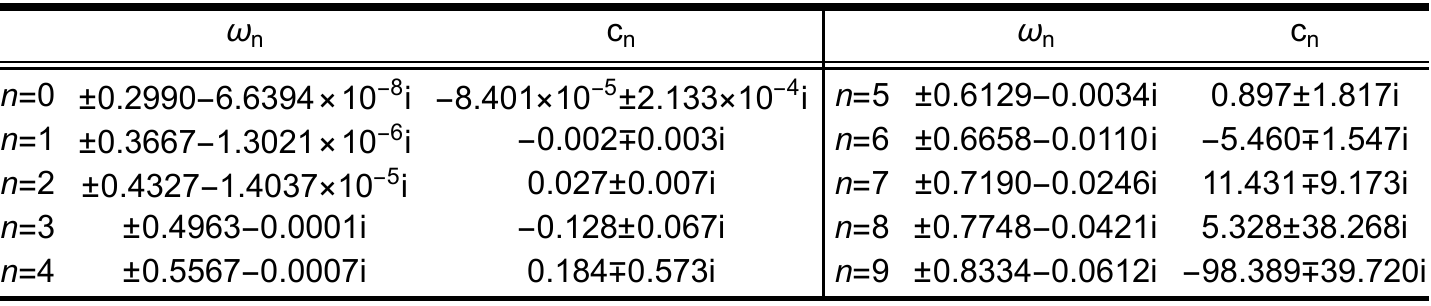} 
\par\end{centering}
\caption{The frequency $\omega_{n}$ and the coefficient $c_{n}$ of the quasinormal
modes of $l=2$ for a scalar field perturbation in the hairy black
hole with $\alpha=0.6$ and $Q=1.0192$. Compared to TABLE \ref{alpha05n15},
fewer quasinormal modes are available to reconstruct the waveform
in FIG. $\ref{alpha06}$.}
\label{alpha06n9}
\end{table}

\subsection{Adjacent Double-peak Potential}

Finally, we consider the hairy black hole solutions with the double-peak
effective potential, where the separation between the peaks $L$ is
comparable to the Compton wavelength of the perturbations. Unlike
the wormhole-like potential, the geometric optics approximation fails,
and hence the time delay between the echoes can be larger than $2L$.
Furthermore, albeit there always exist long-lived modes trapped at
the potential valley, the number of the long-lived modes decreases\ as
$L$ decreases \cite{Guo:2021enm}. In \cite{Guo:2021enm}, we also
found that, near the smaller local maximum of the potential, there
appear sub-long-lived modes, which could play an important role in
the late-time waveform when the contribution from the long-lived modes
is suppressed by their small tunneling rates through the potential
barriers.

\begin{figure}[ptb]
\begin{centering}
\includegraphics[scale=0.7]{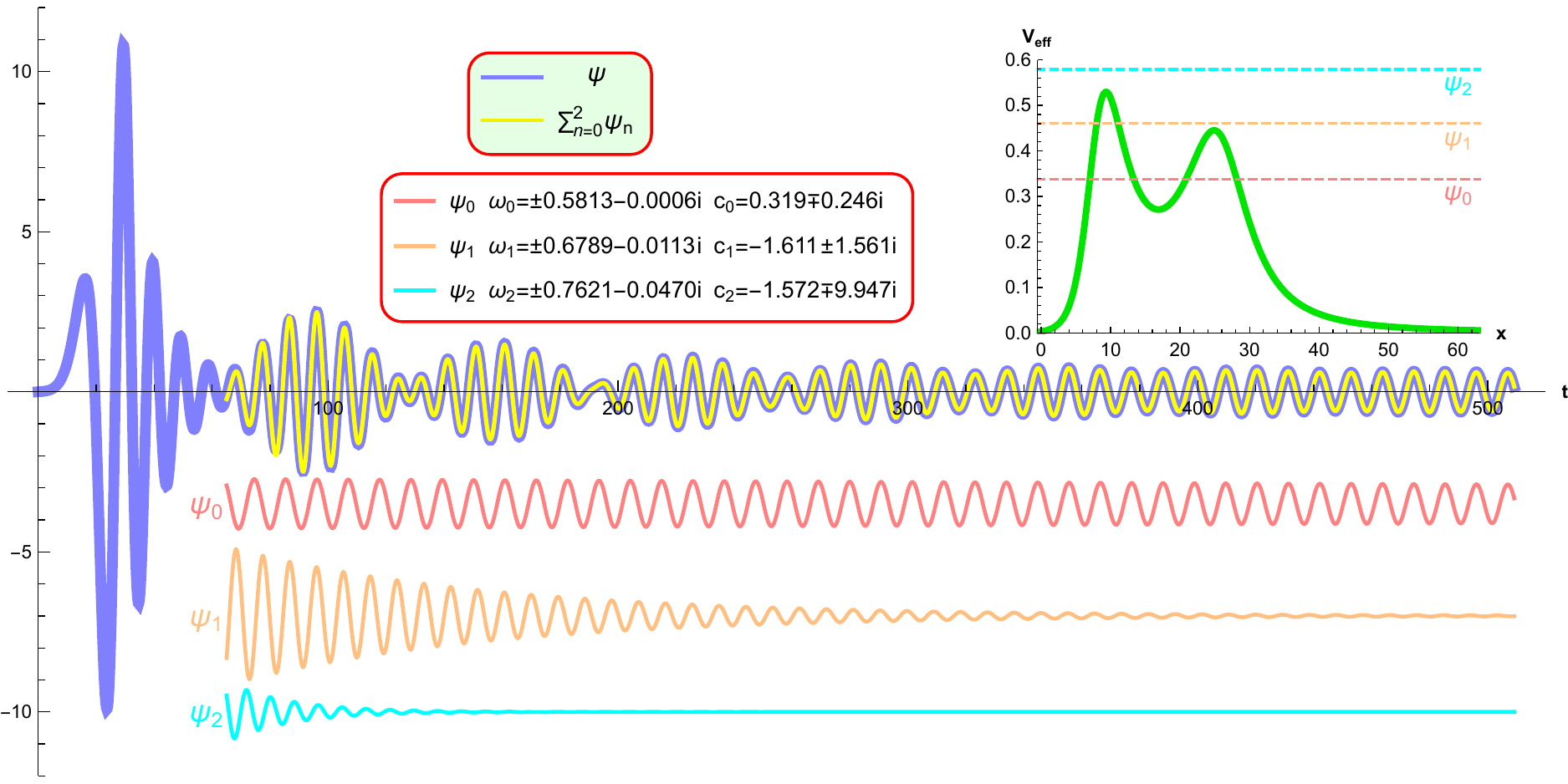} 
\par\end{centering}
\caption{Numerical (blue line) and reconstructed (yellow line) waveforms $\psi$
excited by a Gaussian perturbation near the outer potential peak in
the hairy black hole with $\alpha=0.8$ and $Q=1.0514$. The observer
is far away from the outer peak. Compared to FIGs. $\ref{alpha05}$
and $\ref{alpha06}$, the effective potential has a significantly
shallower and narrower valley. Therefore, only three quasinormal modes,
namely $\psi_{0}$, $\psi_{1}$ and $\psi_{2}$, are needed to reconstruct
the waveform after the primary signal is received. The superposition
of these quasinormal modes leads to the observed echoes. Moreover,
since the geometric optics approximation is invalid, the time delay
between the echoes is obtained from the real parts of the quasinormal
frequencies instead of the distance between the peaks. The dashed
horizontal lines in the inset represent $\left(\operatorname{Re}\omega_{n}\right)^{2}$
of $\psi_{0}$, $\psi_{1}$ and $\psi_{2}$, indicating that $\psi_{0}$
is trapped near the bottom of the potential valley and therefore a
long-lived mode. After $\psi_{1}$ and $\psi_{2}$ are damped away,
the waveform $\psi$ is only determined by the long-lived mode $\psi_{0}$,
and hence has a sinusoid tail instead of echoes. }
\label{alpha08ud}
\end{figure}

In FIG. \ref{alpha08ud}, we investigate the waveform of a scalar
perturbation propagating in the double-peak effective potential of
the hairy black hole with $\alpha=0.8$ and $Q=1.0514$, where the
inner barrier is higher than the outer one. After the primary signal,
one observes three distinct echoes followed by an apparent sinusoid.
Since a shallow potential valley gives rise to fewer quasinormal modes
\cite{Guo:2021enm}, the late-time waveform $\psi$ can be well reconstructed
by only three lowest-lying modes $\psi_{n\leq2}$. Moreover, one can
read off the echo period $T\approx64.38$ from the quasinormal frequencies
$\omega_{0}$, $\omega_{1}$ and $\omega_{2}$ via eqn. $\left(\ref{eq:AP modes}\right)$.
Therefore, the echoes are separated by a distance $T\approx64.38$,
which is larger than $2L\approx31.10$. In the inset, $\left(\operatorname{Re}\omega_{n}\right)^{2}$
of the the quasinormal modes $\psi_{n\leq2}$ are displayed as dashed
horizontal lines. It shows that the $n=0$ quasinormal mode lives
at the bottom of the potential valley and is a long-lived state with
a very small imaginary part of the quasinormal frequency. Additionally,
$\psi_{1}$ is a sub-long-lived mode that lives near the smaller local
maximum of the double-peak potential. The superposition of $\psi_{0}$,
$\psi_{1}$ and $\psi_{2}$ generates the first echo, whereas the
following echoes are mainly determined by $\psi_{0}$ and $\psi_{1}$.
After the echo signals, only the long-lived mode $\psi_{0}$ remains,
which results in a long sinusoid tail. It is worth emphasizing that
the amplitude of the tail is much larger than that in the single-peak
case since the imaginary part of $\omega_{0}$ is roughly $100$ times
smaller.

\begin{figure}[ptb]
\begin{centering}
\includegraphics[scale=0.7]{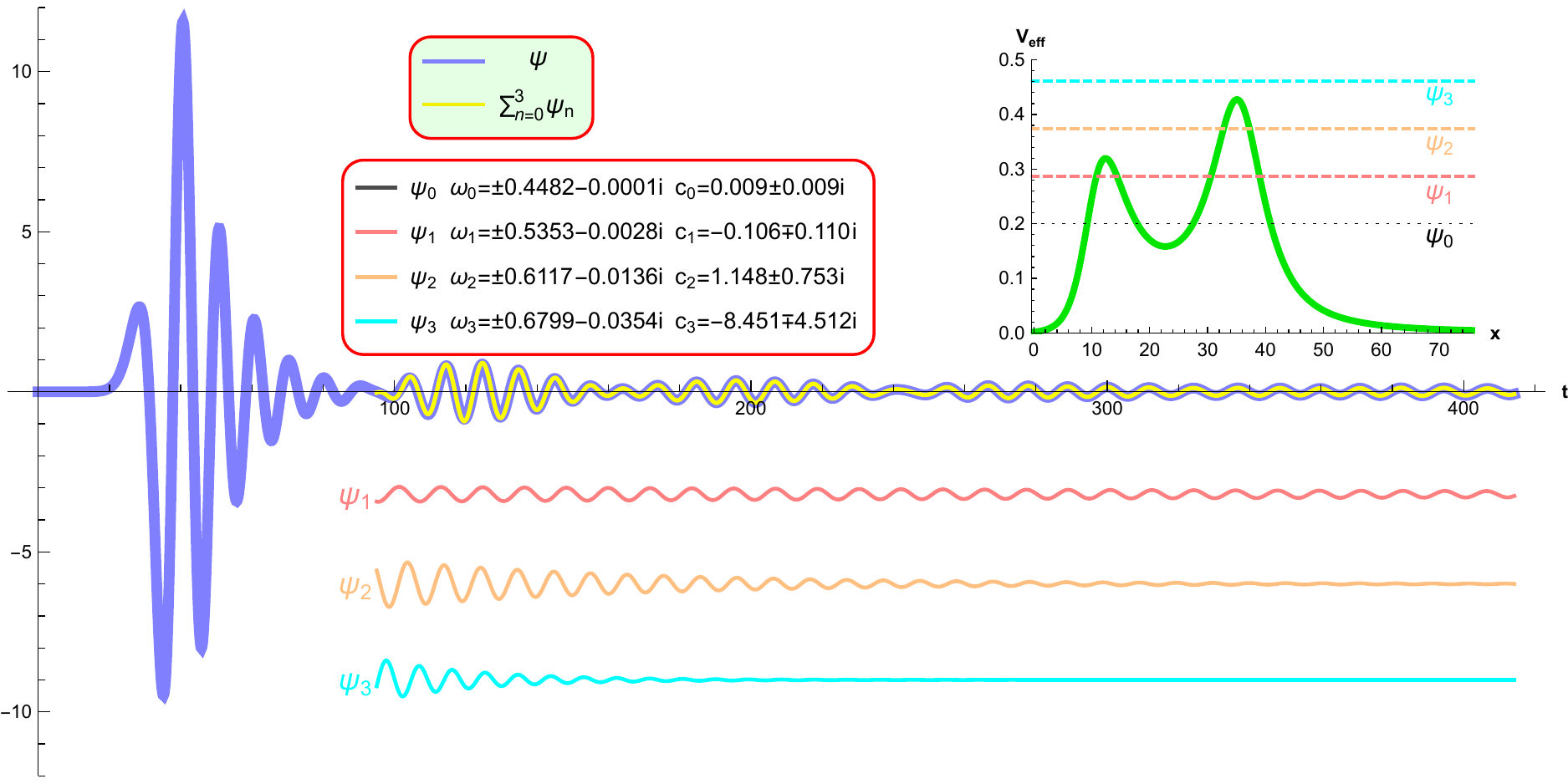} 
\par\end{centering}
\caption{Numerical (blue line) and reconstructed (yellow line) waveforms $\psi$
excited by a Gaussian perturbation near the outer potential peak in
the hairy black hole with $\alpha=0.7$ and $Q=1.0340$, which has
a higher outer potential barrier than in FIG. $\ref{alpha08ud}$.
The higher outer potential barrier reduces the transmission rate of
the perturbation through the barrier. As a result, the long-lived
mode $\psi_{0}$ is negligible, and the sub-long-lived mode $\psi_{1}$
trapped at the smaller local maximum of the potential determines the
late-time sinusoid tail, which has a much smaller amplitude and a
large decay rate than that in FIG. $\ref{alpha08ud}$. In addition,
there appears only one distinguishable echo, which mainly consists
of the quasinormal modes $\psi_{1}$, $\psi_{2}$ and $\psi_{3}$.}
\label{alpha07du}
\end{figure}

To illustrate the effect of the outer potential barrier on the received
signals, we consider the hairy black hole with $\alpha=0.7$ and $Q=1.0340$,
whose effective potential has a higher outer potential barrier than
that in FIG. $\ref{alpha08ud}$. In fact, as the height of the outer
potential barrier increases, perturbations escape from the potential
valley to a distant observer more difficultly. As expected, FIG. $\ref{alpha07du}$
shows that the late time signal detected by a distant observer is
dimmer than that in FIG. $\ref{alpha08ud}$. Specifically, the dashed
horizontal lines in the inset indicate that the quasinormal modes
$\psi_{0}$ and $\psi_{1}$ are long-lived and sub-long-lived modes
trapped at the minimum and the smaller local maximum of the potential,
respectively. Due to the small transmission rate through the high
outer barrier, the long-lived mode $\psi_{0}$ has the negligibly
small coefficient $c_{0}$ and thus contributes little to the late-time
signal. Consequently, the late-time waveform is primarily controlled
by the sub-long-lived mode $\psi_{1}$, which has smaller modulus
of the coefficient $c_{1}$ and decays faster than the long-lived
mode $\psi_{0}$ in FIG. $\ref{alpha08ud}$, hence leading to fewer
echoes and a smaller sinusoid tail.

\section{Conclusions}

\label{sec:Conclusions}

In this paper, we first studied hairy black holes in the EMS model,
where the scalar field is minimally coupled to the gravity sector
and non-minimally coupled to the electromagnetic field. It showed
that the effective potential of scalar perturbations can possess a
single peak or two peaks depending on the black hole parameters. Moreover,
for the double-peak potential, the separation between the peaks can
be significantly larger or comparable to the Compton wavelength of
the perturbations. Considering an initial Gaussian perturbation near
the (outer) potential peak, the evolution of the time-dependent scalar
perturbation was then computed in several hairy black holes to investigate
how the peak structure affects the late-time waveform of the perturbation
received by an observer far away from the (outer) peak. Specifically,
the waveform was obtained by numerically solving the partial differential
equation $\left(\ref{eq:t-x eq}\right)$. To find the frequency content
of the waveform, we also used eqn. $\left(\ref{eq:psi of QNMs}\right)$
to reconstruct the waveform with the associated quasinormal modes.
Our results showed that the numerical and reconstructed waveforms
are in excellent agreement.

After relaxation of the initial perturbation, the observer first detects
a primary signal, which is the reflected wave off the (outer) potential
peak and hence essentially controlled by the quasinormal modes associated
with the (outer) peak. If there is no inner peak, the late-time waveform
after the primary signal is an exponentially decaying sinusoid, which
is the fundamental quasinormal mode. On the other hand, the late-time
waveform in a double-peak potential is mostly determined by the long-lived
and sub-long-lived quasinormal modes, which live near the minimum
and the smaller local maximum of the potential, respectively. When
the distance between the peaks is large, there exist a number of long-lived
and sub-long-lived modes, which produce a train of decaying echo pulses
observed in FIGs. $\ref{alpha05}$ and $\ref{alpha06}$. Remarkably,
if the number of long-lived modes is small enough, echo signals can
disappear for some time and then reappear (see FIG. $\ref{alpha06}$).
When the potential peaks are close enough, there exist only one long-lived
mode and one sub-long-lived mode. The superposition of the long-lived,
sub-long-lived and other low-lying modes give a few observed echoes
following the primary signal (see FIGs. $\ref{alpha08ud}$ and $\ref{alpha07du}$).
For a low outer potential barrier, the long-lived mode dominates the
waveform of the perturbation after other modes are damped away, producing
a very slowly decaying sinusoid tail (see FIG. $\ref{alpha08ud}$).
For a high outer potential barrier, the long-lived mode is suppressed,
and the sub-long-lived is then responsible for the sinusoid tail of
the waveform, which decays faster and has a much smaller amplitude
(see FIG. $\ref{alpha07du}$).

For spherically symmetric hairy black holes, the connection between
double-peak effective potentials and the existence of multiple photon
spheres outside the event horizon has been discussed in \cite{Guo:2021enm}.
The late-time waveform excited by a scalar perturbation may provide
a smoking gun for the detection of black holes with multiple photon
spheres. It will be of great interest if our analysis can be generalized
beyond spherical symmetry and for more general black hole spacetimes.
\begin{acknowledgments}
We are grateful to Yiqian Chen and Xin Jiang for useful discussions
and valuable comments. This work is supported in part by NSFC (Grant
No. 12105191, 11947225 and 11875196). Houwen Wu is supported by the
International Visiting Program for Excellent Young Scholars of Sichuan
University. 
\end{acknowledgments}

 \bibliographystyle{unsrturl}
\bibliography{ref}

\end{document}